\documentclass[twocolumn,showpacs,preprintnumbers,amsmath,amssymb]{revtex4}

\usepackage{graphicx}
\usepackage{dcolumn}
\usepackage{bm}

\begin{document}

\title{Decoherence and dephasing errors caused by D.C. Stark effect in rapid ion transport}

\author{Hoi-Kwan Lau\footnote{kero.lau@utoronto.ca} and Daniel F.V. James\footnote{dfvj@physics.utoronto.ca}}
 \affiliation{ Department of Physics, University of Toronto, 60 Saint George Street, Toronto, M5S 1A7, Ontario, Canada}

\date{\today}

\begin{abstract}
We investigate the error due to D.C. Stark effect for quantum information processing for trapped ion quantum computers using the scalable architecture proposed in J. Res. Natl. Inst. Stan. \textbf{103}, 259 (1998) \cite{Wineland:1998p10591} and Nature \textbf{417}, 709 (2002) \cite{, Kielpinski:2002p7096}.  

As the operation speed increases, dephasing and decoherence due to the D.C. Stark effect becomes prominent as a large electric field is applied for transporting ions rapidly.
We estimate the relative significance of the decoherence and dephasing effects and find that the latter is dominant.  
We find that the minimum possible of dephasing is quadratic in the time of flight, and an inverse cubic in the operational time scale.  From these relations, we obtain the operational speed-range at which the shifts caused by D.C. Stark effect, no matter follow which trajectory the ion is transported, are no longer negligible.
Without phase correction, the maximum speed a qubit can be transferred across a 100 micron-long trap, without excessive error, in about 10 ns for $^{40}\textrm{Ca}^+$ ion and 50 ps for $^{9}\textrm{Be}^+$ ion.  
In practice, the accumulated error is difficult to be tracked and calculated, our work gives an estimation to the range of speed limit imposed by D.C. Stark effect.
\end{abstract}



\maketitle

\section{Introduction}

Since it was realized that quantum algorithms can speed up complicated computational tasks, such as factorizing a large integer, which cannot be efficiently performed by known classical algorithms, building a quantum computer (QC) has become one of the ambitious goals in modern physics \cite{ANielsen:2000p5658, Barnett:2009p8933}.  Among current proposals for physical implementations of a QC, in many ways the ion trap proposal of Cirac and Zoller \cite{Cirac:1995p2238} seems the most auspicious at the moment.  Exploiting well-developed techniques from quantum optics and atomic physics, entanglement of up to 14 ions \cite{Haffner:2005p7378,Monz:2011p10323}, high fidelity gates and readout \cite{Myerson:2008p7428, Leibfried:2003p7149}, and long coherence time quantum memory for more than 10 s \cite{Langer:2005p6869} have been recently demonstrated.  Simple quantum algorithms including Deutsch-Jozsa algorithm \cite{Gulde:2003p7478} and quantum teleportation \cite{Riebe:2004p7558, Barrett:2004p7540}, as well as verification of the Bell inequality \cite{Rowe:2001p7594} have been successfully realized on ion-trap QC.

Nevertheless, we are still far from having a quantum computer with the computing power higher than (or even comparable to) classical computers.  Apart from the small number of qubits entangled, when comparing current capabilities with the resources needed for a practical quantum processor, the speed of quantum operations is another issue that limits the clock rate of an ion-trap QC.  Consider a QC is built to run the Shor algorithm.  It would be of great technical interest if it could break a RSA classical cryptography code in a few days, but it would obviously be even more appealing to break the code, for example, in a few hours.  Then each quantum logic gate has to be performed at the time scale of $\mu$s \cite{Fowler:2004p8100,RSA}.  
When employing fault-tolerant techniques, each quantum logic gate consists of several concatenated rounds of physical quantum operations on ion qubits, including transportation, cooling, and laser interaction \cite{Wineland:1998p10591, Kielpinski:2002p7096}.  
Assessing the true time specifications for a QC in a complicated problem dependent on paradigm choices such as circuit-based versus measurement-based QC, error correction code used, and the implementation of the algorithm to be performed.  
But we can assert quantum computation is more useful and promising if the speed of each physical quantum operation is of nanosecond scale \cite{Stock:2009p3341}.

In this paper, we investigate the effect of D.C. Stark shifts during ion-trap QC operation according to the scalable architecture proposed by Wineland \textit{et al.} \cite{Wineland:1998p10591} and further elaborated by Kielpinski, Monroe and Wineland in \cite{Kielpinski:2002p7096}.  The QC consists of large number of interconnected ion traps.  Ions are stored in memory region, which is far from interaction region where entanglement operations take place.
During quantum computing, ions are moved from the storage region to the interaction region, and transported back to storage region after operations.  The transport of ions can be done by changing electric potential of each trap, so that an effective non-equilibrium electric field is induced.

Fast transport of the ion requires a large electric field which will result in a variety of potentially detrimental effect.  One problem is that a large electric field is less stable, the field fluctuations will heat up the ion to motional excited states \cite{Rowe:2002p6849}.  Cycles of sympathetic cooling are required to bring the ion back to its motional ground state for precise logic operations, but the operation is time-consuming and thus not preferable in high speed quantum computation.  The motional heating effect is anticipated to be reduced by improving experimental techniques, such as using surface traps and coating the electrodes of the trap \cite{Rowe:2002p6849}, or transporting the ions under trajectories with minimal vibrational quanta excitation \cite{Torrontegui:2011p9810}.  Besides, some proposals for entanglement gates remain effective even though the ion has small motional excitation \cite{Srensen:1999p7195, James:2000p10891}. 

More seriously, a large electric field will induce a D.C. Stark effect onto the \textit{internal} electronic states of ions.  Due to the detuning of energy levels and mixing of eigenstates, encoded quantum information will be phase-shifted and excited out of the computational basis.  This issue was first discussed more than 10 years ago by Wineland \textit{et al.} (see ref. \cite{Wineland:1998p10591} p.310); its significance was neglected at that time due to the low operational speed they envisioned.  As technology advances this effect will become important, and a reappraisal is warranted.
The aim of this article is to find out the relation between the total effects by D.C. Stark shift, the size of trap,  trajectory and total time of the flight of ion qubits.  From the relation, we can define the `threshold speed', the speed of ion-shuttling above which the influence of D.C. Stark effect becomes significant.  

Our paper is organized as follows.  In section \ref{sec:motion}, we calculate the wave function of the ion during the shuttling process.  A quantum mechanical treatment is employed, and we find the ion remains in coherent state during the trajectory.  Therefore the classical acceleration of the ion can be obtained, from which the net electric field experienced by the ion is determined.  The total dephasing caused by D.C. Stark shifts is calculated in section \ref{sec:internal}.  
In section \ref{sec:min_phase}, we minimize the dephasing effect with respect to the ion-transport trajectory; a threshold speed is deduced from the relation between minimum phase shift and the time of flight.
In the next section \ref{sec:phase}, we calculate the threshold speed for various choices of ion qubits.  Significance of the decoherence effect is discussed in \ref{sec:decoherence} and we summarize our results in section \ref{sec:summary} with some discussion.  For simplicity, we do not consider the junctions where ions are transferred between different traps; the ion is assumed to move along a straight trap only.  

\section{Motion of ion \label{sec:motion}}

Let us consider an ion is confined in a trap.  Its degrees of motion in the $y$ and $z$ directions are restricted by a strong radio frequency electric field, so the ion is constrained to move only along the $x$ direction.  In current experiments of ion trap QC, the potential of the r.f. trapping field is of the order of 100 V, oscillating at radio frequency at about 100 MHz \cite{Rowe:2002p6849}.  In spite of the strength of the r.f. field, the wave function of the ion is concentrated and peaked at the saddle point of the potential in the $y-z$ plane, on which the electric field vanishes.  Therefore the r.f. field only induces a quadrupolar Stark effect on the ion, which is a higher order effect than the dipole Stark effect during the shuttling and will be neglected in our discussion.  For the same reason, the D.C. Stark effect caused by micromotion of the ion in the $y-z$ plane is also neglected.

In the $x$ direction, the ion is weakly trapped by an electrostatic field.  Close to a minimum of the axial electrostatic potential, the Hamiltonian of the system can be approximated by a harmonic oscillator (Here we only consider harmonic potentials, more general trapping potential and detailed dynamics of trapped ion are discussed in \cite{Torrontegui:2011p9810}).  During shuttling, the strength of the electrostatic field is changed so that the potential well is displaced.  Carefully tuning the strength of the electrostatic field can ensure the strength of harmonic well is constant throughout the process.
Here we assume the electric field is well controlled so the motional heating caused by field fluctuation can be neglected, this assumption will be discussed later.
The time-dependent Hamiltonian is given by the formula
\begin{equation}\label{eq:Hamiltonian}
\hat{H}_M(t) \equiv \frac{\hat{p}^2}{2m}+\frac{1}{2} m \omega^2 \left[\hat{x}-s(t)\right]^2~,
\end{equation}
where $\hat{p}$ is the momentum operator of the ion in $x$ direction; $\omega$ is the harmonic frequency of the potential well formed by electrostatic fields; $s(t)$ is the position of the center of the potential well, which is time dependent.

One can solve for the ion wavefunction $|\Psi(t)\rangle$ governed by the Hamiltonian Eq.~(\ref{eq:Hamiltonian}) in a few steps.  First, the ion should be oscillating around the center of the potential well, so we consider a wavefunction translated by a distance $s(t)$.  This can be done by applying a displacement operator $\hat{D}$ on the state.  We define a new state $|\chi(t)\rangle$ by the formula
\begin{equation}
|\chi(t)\rangle \equiv \hat{D}\left[-\sqrt{\frac{m\omega}{2\hbar}}s(t) \right] |\Psi(t)\rangle~,
\end{equation}
where \cite{Glauber:1963p8906}
\begin{equation}
\hat{D}[v] \equiv \exp(v \hat{a}^\dag-v^\ast \hat{a})~.
\end{equation}
The Schr\"odinger equation becomes
\begin{equation}
i \hbar \partial_t |\chi(t)\rangle = \left(\hat{H}_0 + \dot{s}(t) \hat{p} \right) |\chi(t)\rangle
\end{equation}
where $\hat{H}_0$ is the untranslated harmonic oscillator Hamiltonian, i.e. Eq.~(\ref{eq:Hamiltonian}) with $s(t)=0$.

In the interaction picture, i.e. $ \exp\left(-i \hat{H}_0 t \right)|\tilde{\chi}(t)\rangle \equiv |\chi(t)\rangle$, the wavefunction satisfies the following equation
\begin{equation}\label{eq:HI}
\partial_t |\tilde{\chi}(t)\rangle = \left(\dot{u}(t) \hat{a}-\dot{u}^\ast(t)\hat{a}^\dag \right)|\tilde{\chi}(t)\rangle~,
\end{equation}
where 
$\hat{a}$ and $\hat{a}^\dag$ are lowering and raising operator of $\hat{H}_0$; $u(t)$ is defined as follows:
\begin{equation}
u(t) \equiv -\sqrt{\frac{m \omega}{2 \hbar}}\int_0^t \dot{s}(t') \exp(i \omega t') dt'~.
\end{equation}
The solution of Eq.~(\ref{eq:HI}) is straightforward  \cite{Glauber:1963p8906}:
\begin{equation}
|\tilde{\chi}(t)\rangle = \hat{D}\left[ u(t) \right] e^{\Phi(t)} |\tilde{\chi}(0)\rangle~,
\end{equation}
where $\Phi(t)$ is a time-dependent overall phase of the ion given by
\begin{equation}
\Phi(t) \equiv \frac{1}{2}\int^{t}_0 [u(t') \dot{u}^\ast(t') - u^\ast(t') \dot{u}(t')] dt'~.
\end{equation}
Combining these results, the wavefunction of the ion is therefore given by the formula
\begin{eqnarray}
|\Psi(t)\rangle &=& \hat{D}\left[\sqrt{\frac{m\omega}{2\hbar}}s(t) \right] \exp\left(-i \hat{H}_0 t/\hbar \right) \nonumber \\ &&
\times \hat{D}\left[ u(t) - \sqrt{\frac{m\omega}{2\hbar}}s(0) \right]  e^{\Phi(t)} |\Psi(0)\rangle~.
\end{eqnarray}


If we assume the initial state of ion is the motional ground state of the harmonic oscillator centered at $x=0$, i.e.
$|\Psi(0)\rangle=|0\rangle$,
then the motional state of the ion remains in a coherent state throughout the whole process, i.e. $|\Psi(t)\rangle = |\alpha(t) \rangle$, with the amplitude $\alpha(t)$ given by the formula
\begin{equation}
\alpha(t)\equiv\sqrt{\frac{m \omega}{2 \hbar}} \left(s(t) -e^{-i \omega t} \int^t_0 \dot{s}(t_1) e^{i \omega t_1} dt_1  \right).
\end{equation}
The expectation value of position $q(t)$ is hence
\begin{equation}\label{eq:rhos}
q(t)=s(t)-\int^t_0 \dot{s}(t_1) \cos \left[ \omega (t-t_1) \right]dt_1 ~.
\end{equation}
Because of the classical nature of coherent states, the net electric field, $\vec{\xi}=\xi \hat{e}_x$ where $\hat{e}_x$ is the unit vector along $x$ direction, experienced by the ion is related to the acceleration of the expectation value of position by the Newton's third law, viz.,
\begin{eqnarray}\label{eq:e-field}
\xi(t) &=& \frac{m}{e} \ddot{q}(t) \nonumber \\ &=&\frac{m\omega^2}{e} \int^t_0 \dot{s}(t_1) \cos [\omega (t_1-t)]dt_1~.\label{eq:e-field}
\end{eqnarray}

\section{Phase shift due to D.C. Stark effect\label{sec:internal}}

In this section, we consider the total phase shifted during the shuttling process.  The effect of the applied electric field on the internal structure of the ion will be described by an additional Hamiltonian $\hat{H}_{\textrm{Stark}}=-\vec{d} \cdot \vec{\xi}$, where $\vec{d}$ is the dipole operator. Since the fastest shuttling time achievable in foreseeable future would not be faster than 10 ps, which is about four order slower than the reciprocal of the atomic transition frequency, we can obtain the D.C. Stark shift energy at time $t$ by the time independent perturbation theory \cite{book:Liboff}, viz., 
\begin{equation}\label{eq:perturb}
E_n(t) \approx E^{(0)}_n +E^{(2)}_n(t);~E^{(2)}_n(t)= \sum_{m\neq n} \frac{|e\xi(t) \langle m|\hat{x}|n\rangle|^2}{\hbar \omega_{nm}}
\end{equation}
where electric field points to $z$ direction only; $\hbar\omega_{nm}=\hbar\omega_n - \hbar \omega_m$ is the energy difference between energy eigenstates $|n\rangle$ and $|m\rangle$ of the unperturbed Hamiltonian.


Suppose two unperturbed energy eigenstates $|f\rangle$ and $|i\rangle$ are chosen as the computational basis of the ion qubit.  We work in the interaction picture.  The ion is initially encoded with some quantum information as the state $\alpha|i\rangle+\beta|f\rangle$.  Because the D.C. Stark energy of each computational basis states are different, the qubit state gains an extra phase after shuttling,
\begin{equation}
\alpha|i\rangle+\beta|f\rangle \rightarrow \alpha|i\rangle+\beta e^{i\phi}|f\rangle~, \nonumber
\end{equation}
up to an unimportant global phase.  Substituting from Eq.~(\ref{eq:e-field}) and (\ref{eq:perturb}), the extra phase factor $\phi$ is given by
\begin{equation}\label{eq:phase}
\phi = \frac{m^2}{\hbar} \left(\sum_{m\neq i} \frac{|\langle m|\hat{x}|i\rangle|^2}{\hbar \omega_{im}} - \sum_{m\neq f} \frac{|\langle m|\hat{x}|f\rangle|^2}{\hbar \omega_{fm}} \right)\zeta[q(t)]~,
\end{equation}
where
\begin{equation}\label{eq:zeta}
\zeta[q(t)] = \int^T_0 \ddot{q}(t)^2 dt~.
\end{equation}
All terms except $\zeta$ depend on atomic structure of the ion only, so a good choice of ion qubit would give a small magnitude of these terms; $\zeta$ is a functional of the trajectory of the ion $q(t)$ which is the same for any choice of ions and any computational basis states being used.

\section{Minimum possible phase shift\label{sec:min_phase}}

From Eq.~(\ref{eq:phase}) and (\ref{eq:zeta}), we see that the total phase shift due to the D.C. stark effect is determined by the trajectory.  
Among all the trajectories the ion travel, there is one, the optimal trajectory $q_0$, which produces the minimum phase shift, $\phi_{\textrm{min}}(L,T)$, depending on the length $L$ and time $T$ of the flight.  When the ion is shuttled in a particular time scale at which $\phi_{\textrm{min}}$ becomes significant, the phase shift has to be corrected by some measures, either unitaries or quantum error-correcton-codes (QECC), regardless the trajectory's form and whether it is predictable or not.  This thus places a threshold on ion-shuttling speed.

To find the minimum possible phase $\phi_{\textrm{min}}$, we have to find an ion trajectory $q(t)$, for which $\zeta$ is a minimum.  We solve this problem using calculus of variation (see, e.g. \cite{IlichAkhiezer:1962p8200}).  By defining 
\begin{equation}\label{eq:Lagrange_constraint}
\gamma(t)=\dot{q}(t), 
\end{equation}
minimizing $\zeta$ is equivalent to minimizing $\tilde{\zeta}$, defined as
\begin{equation}
\tilde{\zeta}[q,\dot{q},\gamma,\dot{\gamma},t] = \int^T_0 \left\{\dot{\gamma}^2(t)+\mu(t) [\dot{q}(t)-\gamma(t)]\right\}dt~, 
\end{equation}
where $\mu(t)$ is the Lagrange multiplier corresponding to our assumption Eq.~(\ref{eq:Lagrange_constraint}).  It is beneficial to minimize $\tilde{\zeta}$ because it is a functional of $q$, $\gamma$, and their first order time derivative only (i.e. no high order derivatives).  Then we can easily write down the Euler equation with respect to $q$ and $\gamma$
\begin{eqnarray}
\mu(t)-2\ddot{\gamma}(t)&=&0 \\
\dot{\mu}(t)&=&0~.
\end{eqnarray}
Incorporating these equations and the constraint Eq.~(\ref{eq:Lagrange_constraint}), we find $q_0$ obeys
\begin{equation}\label{eq:deom}
\ddddot{q_0}(t)=0~.
\end{equation}

We assume the ion is displaced by distance $L$ after the process, the initial and final position can be set as $q(0)=0$ and $q(T)=L$.  Additionally, we require the ion remains in the motional ground state before and after the shuttling, so the initial and final velocity both vanish, i.e. $\dot{q}(0)=\dot{q}(T)=0$.  Putting these initial conditions into Eq.~(\ref{eq:deom}), we find the optimal trajectory is
\begin{equation}\label{eq:eom}
q_0(t) = L\left(-2\frac{t^3}{T^3}+3\frac{t^2}{T^2}\right)~,
\end{equation}
which gives $\zeta(q_0)=12 L^2/T^3$.  Hence the minimum phase shift is
\begin{equation}\label{eq:min phase}
\phi_{\textrm{min}} = \frac{12 m^2 L^2}{\hbar T^3} \left(\sum_{m\neq i} \frac{|\langle m|\hat{x}|i\rangle|^2}{\hbar \omega_{im}} - \sum_{m\neq f} \frac{|\langle m|\hat{x}|f\rangle|^2}{\hbar \omega_{fm}} \right)~.
\end{equation}

\begin{figure}
\begin{center}
		\includegraphics{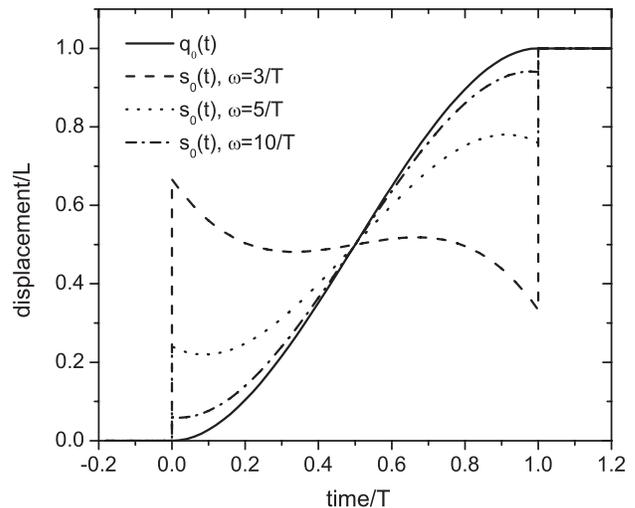}
	\end{center}
	\caption{\label{fig:trajectory}Time dependence of optimal trajectory of ion $q_0(t)$, and optimal trajectory of potential well $s_0(t)$ for different trapping frequencies.}
\end{figure}

We note that the ion does not always stay at the equilibrium position of the potential well $s(t)$.  As expected by classical intuition, the ion keeps sloshing as the well is displaced.  Transporting the ion in a desired trajectory has to be achieved by tuning the trapping electric field carefully to balance this sloshing.  The position of the well can be obtained by combining Eq.~(\ref{eq:rhos}) with its second derivative.  The displacement of the well $s_0(t)$ which produces the optimal trajectory of ion is given by the following expression:
\begin{eqnarray}
s_0(t) &=& q_0(t)+\frac{\ddot{q}_o(t)}{\omega^2}\nonumber \\&=&L\left(-2\frac{t^3}{T^3}+3\frac{t^2}{T^2}-12\frac{t}{\omega^2 T^3}+\frac{6}{(\omega T)^2}\right)~.
\end{eqnarray}
During the shuttling, i.e. $0<t<T$, and the ion should stay at the equilibrium point of the well before and after the travel, i.e. $s(t<0)=0$ and $s(t>T)=L$.  Both $q_0(t)$ and $s_0(t)$ are shown schematically in Fig.~\ref{fig:trajectory}.  

We can see the position of the potential well jumps sharply at $t=0$ and $t=T$.  This means that the trapping electric fields are sharply changed at the beginning and at the end.  However, the sudden variation of electric field does not cause a significant D.C. Stark effect.  It is because the energy perturbation only depends on the strength of electric field but not its time derivative within the dipole approximation.  

We have found the optimal trajectory which the phase shift is minimum, but the motion of ion cannot be exactly controlled or predicted in practice, and trajectories other than $q_0(t)$ may be employed due to experimental convenience.  To verify the appropriateness of our threshold speed deduced from $\phi_{\textrm{min}}$, we have to see if the phase shift would vastly increase if the ion takes other trajectories.  Firstly we consider a more experimental realizable trajectory which the electric field is tuned gradually, i.e. no sharp slips of potential well.  We try the trajectory described by
\begin{equation}\label{eq:5path}
q(t)=L(6t^5/T^5-15t^4T^4+10t^3/T^3)~, 
\end{equation}
where the position of the electric potential well varies continuously as time, even at the beginning and the end.  We find for this trajectory that $\zeta=17.1 L^2/T^3$, which is larger than $\zeta(q_0)$ but remains in the same order of magnitude.  


Secondly, we consider the trajectory of the experimental setting of Rowe \textit{et al.} \cite{Rowe:2001p7594}, where the location of the trapping potential $s_R$ varies as
\begin{equation}\label{eq:sRowe}
s_R(t) = L \sin^2\left(\frac{\pi t}{2T}\right)~,
\end{equation}
and the frequency of the potential well is $\omega_R = 2\pi \times 2.9~\textrm{MHz}$.  The trajectory of the ion $q_R$ can be obtained from Eq.~(\ref{eq:rhos}), and the value of trajectory dependent term $\zeta$ in the expression of phase shift Eq.~(\ref{eq:phase}) is $\zeta(q_R) = 24.3 L^2/T^3$, therefore the dephasing is double of $\phi_{\textrm{min}}$.  For these two examples, the phase shifts are at the same order as $\phi_{\textrm{min}}$, therefore the threshold speed derived from $\phi_{\textrm{min}}$ is a useful benchmark.


\section{Threshold speed of transporting ion qubits\label{sec:phase}}

In this section, we investigate the threshold speed of common ion qubit candidtates.  We consider what if the phase shift due to D.C. Stark effect is neglected, i.e. no unitary correction is conducted.  It will be a problem if its magnitude is to large to be corrected by QECC of the system.  The threshold error rate of the best known QECC now is about 1$\%$ (e.g. \cite{Raussendorf:2007p2845}).  To estimate the order of magnitude of the minimum shuttling time, we let the upper limit of the optimal phase in Eq.~(\ref{eq:min phase}) as $\pi/100$.  
We assume the length of the trap being travelled is $L= 100 \mu m$, which matches setting of current experimental setup \cite{Rowe:2001p7594}.  Two popular proposals of ion qubits are investigated.  One proposal uses electronic states of $^{40}\textrm{Ca}^+$ ion \cite{Riebe:2004p7558} as qubits levels.  The computational basis states are $|i\rangle = |S_{1/2},M_J=-1/2\rangle$ ground state and the $|f\rangle=|D_{5/2}, M_J=-1/2\rangle$ metastable state, which the $T_1$ time is about 1 s.  Another proposal uses hyperfine states of $^9\textrm{Be}^+$ ion \cite{Langer:2005p6869,Barrett:2004p7540,Rowe:2001p7594}.  The computational states we consider are $|i\rangle=|F=2, m=0\rangle$ and $|f\rangle=|F=1,M_F=1\rangle$.  The $T_2$ time is greater than 10 s when the beryllium ion is subjected to magnetic field with strength 0.01194 T \cite{Langer:2005p6869}, while the $T_1$ time for these hyperfine states is longer than centuries \cite{Fuhr:2010p8878}.

Now we calculate the term inside the bracket in Eq.~(\ref{eq:min phase}).  Values of the matrix elements can be calculated from tabulated parameters found in ref. \cite{website:ASD}.  We calculate the term belonging to $|i\rangle$ while the one of $|f\rangle$ can be calculated by the same method.  Let the state $|i\rangle$ belongs to energy level $k$, it can be expressed as a superposition of states with definite magnetic quantum number $|r\rangle$,
\begin{equation}
|i\rangle = \sum_{r\in k} A_{i r} |r\rangle~.
\end{equation}
We pick the intermediate states $|m\rangle$ as energy eigenstates with definite magnetic quantum number, so the summation of $|m\rangle$ involves summing of states in a particular energy level $l$ and then sum over all energy levels.  The transition energy $\hbar \omega_{im}$ is the energy difference between levels $k$ and $l$ and given explicitly in \cite{website:ASD}.  For the matrix elements of dipole operator, we extract the reduced matrix elements,
\begin{equation}
\sum_{m \in l}|\langle m|\hat{x}|i\rangle|^2 = |\langle k || \hat{Q}^1 || l \rangle|^2 \sum_{r\in k}\sum_{m\in l} |A_{i r}|^2 |C_{mr}|^2~,
\end{equation}
where $\hat{Q}^1$ is the rank 1 irreducible tensor operator, $|\langle k || \hat{Q}^1 || l \rangle|^2$ is the reduced density matrix between energy levels $l$, $C_{mp}$ is the Clebsch–Gordan coefficients between $|m\rangle$ and $|p\rangle$ \cite{CG}.  The reduced density matrix can be calculated from the line strength $S_{lk}$ between energy levels $l$ and $k$ provided in \cite{website:ASD}.  By definition,
\begin{equation}
S_{lk} =g_k \sum_{m \in l} \langle p | \vec{d} | m \rangle \cdot \langle m | \vec{d} | p \rangle = e^2 g_k |\langle k || \hat{Q}^1 || l \rangle|^2~,
\end{equation}
where $g_k$ is the degeneracy of the level $k$.  The last relation is derived from the orthonormality condition of Clebsch-Gordan coefficients.  

For the calcium qubit, 
only the states $|m\rangle=|P_{3/2},M_J=-1/2\rangle$ and $|P_{1/2},M_J=-1/2\rangle$ can yield non-zero dipole matrix elements.  By using the formalism mentioned above, we find the minimum phase is 
\begin{equation}\label{eq:phiCa}
\phi^{\textrm{Ca}}_{\textrm{min}}=9.86 \times 10^{-18} \frac{[L^2]}{[T^3]}~,
\end{equation}
where the square bracket denotes the value of quantities in S.I. unit.  Both computational states of beryllium qubit are hyperfine states of the ground electronic state, so $|m\rangle$ are states on $P_{1/2}$ and $P_{3/2}$ levels only.  We find the minimum phase is
\begin{equation}\label{eq:phiBe}
\phi^{\textrm{Be}}_{\textrm{min}} = 2.6 \times 10^{-25} \frac{[L^2]}{[T^3]}~.
\end{equation}

For $L=100~\mu m$ and $\phi_{\textrm{min}} \lesssim \pi/100$, the minimum time of flight for ion qubits are
\begin{eqnarray}
T^{\textrm{Ca}}_{\textrm{min}} &\gtrsim& 14.6~ns \label{eq:TCa}\\
T^{\textrm{Be}}_{\textrm{min}} &\gtrsim& 0.044~ns \label{eq:TBe}~.
\end{eqnarray}

$T^\textrm{Be}_\textrm{min}$ is 3 order smaller than $T^\textrm{Ca}_{\textrm{min}}$ because the Zeeman energy between two qubit states, $E_Z$, is small compared to energy difference between atomic energy levels, $E_A$.  If no magnetic filed is applied and the tiny hyperfine splitting is neglected, the value of the bracket in Eq.~(\ref{eq:min phase}) vanishes since both qubit states are atomic ground states.  The first non-vanishing term would be the first order term of $E_Z/E_A\approx 10^{-6}$, and so the phase change of beryllium qubit is much smaller than calcium qubit. 

\section{Decoherence \label{sec:decoherence}}

Apart from dephasing effect, the electric field also excites the electron into states outside the computational basis states; neglecting this effect will cause decoherence error on quantum information.  We start by writing the D.C. Stark effect Hamiltonian in the interaction picture
\begin{equation}
\hat{V}_I(t) = \frac{m}{e}V_{rn} \ddot{q}(t) e^{i \omega_{rn}t} |r\rangle \langle n|~,
\end{equation}
where 
\begin{equation}
\langle r | \hat{H}_{\textrm{Stark}}|n\rangle = \frac{m}{e} V_{rn} \ddot{q}(t) = -\frac{m}{e} \langle r | \vec{d} \cdot \hat{x}|n\rangle \ddot{q}(t)~.
\end{equation}
Using time dependent perturbation theory, the first three terms of the Dyson series of the propagator $\hat{U}_I(t)$ in the interaction picture are as follows:
\begin{widetext}
\begin{equation}\label{eq:propagator}
\hat{U}_I(t)\approx \hat{I} -\frac{i}{\hbar}\frac{m}{e}V_{rn}\int^t_0 \ddot{q}(t')e^{i \omega_{rn}t'}dt' |r\rangle \langle n| 
-\frac{1}{\hbar^2}\frac{m^2}{e^2} \sum_{r'} V_{rr'}V_{r'n} \int^t_0 \int^{t'}_0 \ddot{q}(t'') \ddot{q}(t')
e^{i (\omega_{r'n}t'' + \omega_{rr'}t')}dt'' dt' |r\rangle \langle n| ~.
\end{equation}
\end{widetext}
The degree of decoherence is determined by the amplitude of non-computational states at $t=T$.  The two most significant terms are: the first order term with $\omega_r\neq \omega_n$, because the states in the same energy level have the same parity and so $V_{nn}=0$; and the second order term with $\omega_r = \omega_n$ such that $\omega_{rr'}+\omega_{r'n}=0$, otherwise there will be a fast oscillating term in the integral and the contribution of the term is reduced by an order of perturbation.

Although the time integrals cannot be solved without the exact form of $q(t)$, we can compare the significance of both terms by estimating their magnitude.  After simple integration by parts, the integral in the first order term will be 
\begin{eqnarray}\label{eq:decoherence1}
\int^T_0 \ddot{q}(t)e^{i \omega_{rn}t}dt &=& \frac{-i}{\omega_{rn}} \Big( \ddot{q}(T) e^{i \omega_{rn}T} - \ddot{q}(0) \Big) \nonumber \\
&&+\frac{i}{\omega_{rn}}\int^T_0 \dddot{q}(t)e^{i \omega_{rn}t}dt
~.
\end{eqnarray}
Since the typical order of atomic transition frequency $\omega_{rn}$ is $10^{15}~s^{-1}$ and the timescale of ion transport is about $10^{-9}$ s, the last term on the right hand side can be neglected as it is $10^{-6}$ smaller than the one in the left hand side \cite{book:Liboff}.  As we are only interested in the order of magnitude, we consider the bracket in the right hand side has roughly the same magnitude as $L/T^2$.  

For the second order term, we can do a trick to estimate its value.  Consider the contribution for $|r\rangle = |n\rangle$.  This corresponds to the phase shift that we have calculated in previous sections.  Therefore we know the time integral can be approximated by $i\zeta(q) /\omega_{r'n}$, which the value is at the order of $10 L^2/T^3$.  We consider $V_{r'n}$ of low lying energy states dominate, and the matrix elements of the states in the same energy levels are in the same order.  The ratio of the first and second order term is estimated by
\begin{equation}
\frac{\textrm{first order}}{\textrm{second order}} \approx 
 \frac{\hbar e T}{10 |V_{rn}| L}~.
\end{equation}
For a qubit moving in a 100 $\mu$m trap within 10 ns, the above ratio is about $10^{-4}$, which means that the second order term should dominate the decoherence effect.

The magnitude of the decoherence error is defined by the fidelity of the quantum state after passing through an error channel.  Consider the D.C. Stark effect has excited the electron to non-computational basis state $|b_i\rangle$,
\begin{equation}
|\Psi\rangle \rightarrow \sqrt{1-\sum_i |\epsilon_i|^2} |\Psi\rangle + \sum_i \epsilon_i |b_i\rangle~,
\end{equation}
so the decoherence error is given by $\sum_i |\epsilon_i|^2$.  For a calcium qubit, the amplitude of the second order term in Eq.~(\ref{eq:propagator}) is on the same order as the phase shift $\phi$, so the magnitude of decoherence error is about $\phi^2$.  If the calcium ion is transported in a speed such that the dephasing error is $10^{-2}$, the decoherence error will become $10^{-4}-10^{-3}$.  Therefore the threshold speed of trapped calcium ion QC is determined by the dephasing effect, i.e. Eq.~(\ref{eq:TCa}).

The second order decoherence term of beryllium qubit vanishes if hyperfine splitting is neglected.  It is because both computational states consist of one orbital state, $|l=0,m_l=0\rangle$, only, i.e.
\begin{eqnarray}
|2,0\rangle_F &=& \left(\frac{1}{\sqrt{2}}|\frac{3}{2},\frac{-1}{2}\rangle_I |\frac{1}{2},\frac{1}{2}\rangle_S +\frac{1}{\sqrt{2}}|\frac{3}{2},\frac{1}{2}\rangle_I |\frac{1}{2},\frac{-1}{2}\rangle_S \right) \nonumber \\
&&\otimes |l=0,m_l=0\rangle \\
|1,1\rangle_F &=& \left(\frac{\sqrt{3}}{2}|\frac{3}{2},\frac{3}{2}\rangle_I |\frac{1}{2},\frac{-1}{2}\rangle_S -\frac{1}{2}|\frac{3}{2},\frac{1}{2}\rangle_I |\frac{1}{2},\frac{1}{2}\rangle_S \right) \nonumber \\
&&\otimes |l=0,m_l=0\rangle~,
\end{eqnarray}
where the first and second number of the states denote their angular momentum quantum number and magnetic quantum number; the subscripts specify the type of angular momentum.  Since the orbital state has no degeneracy, D.C. Stark effect can only produce a phase shift.  Without hyperfine shift, the phase shift on each component of the computational basis is the same, so there is no decoherence.  

Just as the dephasing effect, the second order decoherence term of beryllium qubit is non-vanishing only if hyperfine splitting is included, but its magnitude is reduced by a factor of $E_Z/E_A\approx 10^{-6}$.  Then according to the previous analysis, the first order decoherence term may become dominant.  However, the first order decoherence term can be suppressed by tuning the trajectory of the ion qubit.  Recall we have assumed the bracket in Eq.~(\ref{eq:decoherence1}) takes the value $L/T^2$, which determines the order of magnitude of the decoherence.  However, if we choose a trajectory that the initial and final acceleration of the ion qubit are both zero, then the bracket vanishes.  Hence the first order decoherence effect will be reduced by a factor of $(\omega_{rn}T)^{-1}$, which is a part in $10^6$ for $T=1$ ns.  Although the optimal trajectory $q_0(t)$ does not satisfy this criteria, as shown in Fig.~\ref{fig:trajectory}, we can slightly modify $q_0(t)$.  An example to doing this is to increase the acceleration from zero to $\ddot{q}_0(0)$ initially and decrease from $\ddot{q}_0(T)$ to 0 at the end of the flight, both are conducted slowly comparing to $\omega_{rn}$ but fast comparing to $T$, and the rest of the time follows $q_0(t)$.  We find such modification does not alter the phase shift much, with only a factor of order 1, but the first order decoherence factor is greatly suppressed.

So the second order decoherence term of beryllium qubit remains dominant.  A similar situation pertains to the calcium qubit.  Although the amplitude of non-computational states is on the same order as the dephasing of computational states, the decoherence error is the square of the amplitude.  Therefore, we can state with some evidence that threshold speed of beryllium qubit is determined by the dephasing error, which is smaller than the order of 100 ps as given in Eq.~(\ref{eq:TBe}).

\section{Summary\label{sec:summary}}

In this paper, we have estimated the magnitude of D.C. Stark shifts on quantum information during ion shuttling.
A larger electric field should be employed to transport ion qubits faster for rapid QC operation, but it also induces more serious D.C. Stark effect on the computational basis states, which shifts the relative phase between qubit states and excits the electron out of the computational basis.  Magnitude of these effects are characterized by the trajectory of ion transportation, so the threshold speed can be estmiated by minimizing the trajectory.
We find that dephasing is the most important source of error, its magnitude scales as square of the size of traps and inversely cubic as the operational timescale.  For a trapped-ion QC with characteristics length about 100 $\mu$m, without correction the shortest time for each shuttling $T_{\textrm{min}}$ is about 10 ns if $^{40}\textrm{Ca}^+$ ion qubit is employed.  
On the other hand, using hyperfine states of $^9\textrm{Be}^+$ ion as qubit states will be more error-resistant.  We find that the total phase shift due to D.C. Stark effect is 6 order smaller than the implementation using calcium ion.  Beryllium ion can be transported as fast as 50 ps before dephasing and decoherence effects become significant.

In principle, the average Stark shift might be assessed by Ramsey interferometry, and then corrected by unitary transformations.  In practice, however, the ion is undergoing a complex trajectory involving acceleration and deceleration, moving in straight lines, turning around bends and through junctions, disengaging the individual ion from the storage register and the logic trap; all of these effects will be too complicated to track and calculate the Stark shift accurately.  The uncertainties will turn the shifts into dephasing and decoherence \textit{errors}.  Error correction will perforce be needed, and the requirements of fault-tolerant quantum computing (in particular, ensuring the error be below some threshold) will place a speed limit on the operation of the QC.  Magnitudes of the errors depend on the setup of the ion-trap system.  In any case, our result is a useful reference to the speed limit.  As a rough illustration, suppose the overall Stark shifts can be evaluated, with accurately tracking the trajectory, well-controlled electric field, and other very precise experimental techniques, up to $90\%$ accuracy for a particular ion-trap QC.  The speed limit of this setup, due to the $10\%$ uncertainty, can be evaluated by Eq.~(\ref{eq:phiCa}) and (\ref{eq:phiBe}) as roughly two times of the threshold speed.




A possible way to reduce this lower bound is to reduce the size of ion trap.  But this method is inefficient because $T_{\textrm{min}}$ only scales at $L^{2/3}$, and the reduction of size by half already require tremendous advance in engineering techniques.  Another method is to encode quantum information onto the decoherence free subspace of two calcium ions \cite{Kielpinski:2002p7096,Roos:2006p8800}, e.g. $|0\rangle \rightarrow |i_1f_2\rangle$, $|1\rangle \rightarrow |f_1i_2\rangle$.  As long as the traveled trajectory by both ions are approximately the same, the total dephasing effect on the logical qubit can be reduced by several orders  \cite{Kielpinski:2002p7096}.  
On the other hand, there are alternative scalable ion-trap QC architecture which requires much fewer times (the measurement-based ion-trap QC, c.f. \cite{Stock:2009p3341,Raussendorf:2001p3202}), or even no ion-shuttling (the ion-photon network model, c.f. \cite{Blatt:2008p5155, Moehring:2007p10840, Duan:2010p5182}); so they are not seriously affected by the D.C. Stark effect.


We thank the useful comments from the anonymous referee.  We would like to acknowledge support from the NSERC CREATE Training Program in Nanoscience and Nanotechnology.



\bibliographystyle{phaip}
\pagestyle{plain}
\bibliography{Stark}

\end{document}